\documentclass{ws-ijmpa}
\usepackage{graphics}
\usepackage{amssymb}
\usepackage{amsfonts}
\usepackage{amsmath}
\usepackage{pstricks}
\usepackage[T2A]{fontenc}
\def\lsim{\mathrel{\rlap{
\lower4pt\hbox{\hskip-3pt$\sim$}}
\raise1pt\hbox{$<$}}}     
\def\gsim{\mathrel{\rlap{
\lower4pt\hbox{\hskip-3pt$\sim$}}
\raise1pt\hbox{$>$}}}     

\begin{document}

\markboth{}
{Effects of model parameters in thermodynamics of the PNJL model}

%
\catchline{}{}{}{}{}
%

\title{EFFECTS OF MODEL PARAMETERS IN THERMODYNAMICS OF THE PNJL MODEL}

\author{FRIESEN A.V.}

\address{Bogoliubov Laboratory of Theoretical Physics, Joint Institute for
Nuclear Research, 141980 Dubna, Russia\\
avfriesen@theor.jinr.ru}

\author{KALINOVSKY Yu.L.}

\address{Laboratory of Information Technologies, Joint Institute for Nuclear
Research, 141980 Dubna, Russia \\
Higher Mathematics Department, University "Dubna", Dubna, Russia \\
kalinov@jinr.ru}

\author{TONEEV V.D.}

\address{Bogoliubov Laboratory of Theoretical Physics, Joint Institute for
Nuclear Research, 141980 Dubna, Russia\\
toneev@theor.jinr.ru}

\maketitle

\begin{history}
\received{Day Month Year}
\revised{Day Month Year}
\end{history}

\begin{abstract}
The thermodynamic behavior of the two-flavor($N_f=$2) three-color ($N_c=3$)
Polyakov-loop-extended Nambu-Jona-Lasinio model at the finite
chemical potential is investigated. New lattice gluon data for 
gluon thermodynamics are
used defining the effective potential within polynomial and
logarithmic forms of its approximation. We study the effects of
using different sets of data and different forms of the potential
on thermodynamic properties of hot and dense matter. It is
found that the PNJL thermodynamics depends  stronger on the form
of the effective potential than on the used lattice data set.
Particular attention is paid to the phase diagram in the $(T,\mu)$
plane.

\end{abstract}

\ccode{PACS numbers:  11.30.Rd, 12.20.Ds, 14.40.Be}

\section{Introduction}

In the last years the   Polyakov-loop-extended Nambu--Jona-Lasinio
(PNJL) model~\cite{MMO02}\cdash\cite{GMMR06} was widely used
in the study of thermodynamics and the phase diagram of hot
and dense matter. Results of this research are expected to play an
important role in our understanding of the evolution of the early
universe and physics of heavy ion collisions at relativistic
energies. This improved field theoretical model is fundamental for
interpreting the lattice QCD data and extrapolating into regions
not yet accessible for lattice simulations.

An attractive property of the PNJL model is the synthesis of the
Polyakov loop dynamics with the Nambu--Jona-Lasinio model,
combining the two principal nonperturbative features of low-energy
QCD: confinement and spontaneous chiral symmetry breaking. A
particular feature of this model is that one can uniquely determine
the coupling between the chiral condensate, which is an order
parameter of the chiral phase transition when $m_q\to$0, and the
Polyakov loop, which is the order parameter for the deconfinement
phase transition in the limit $m_q\to\infty$. The model is remarkably
successful in reproducing lattice data on the QCD
thermodynamics~\cite{ratti,rosner}\cdash\cite{zhang}.

However, the choice of the parameter set as well as regularization
of integrals, as noted by several authors,  is a nontrivial question.
As it is well known, the order of the phase transition in the
$(T,\mu)$ plane is sensitive to the parameter choice. It was
already noted~\cite{CRSK04,buballa} that one can choose different
sets of parameters which allow for a first order phase
transition, giving a reasonable fit to physics observables in the
vacuum but predicting different physical scenarios at finite
temperature $T$ and chemical potential $\mu$. In addition,
recently new lattice data for the pure gluon QCD sector defining
the effective Polyakov loop potential have been
obtained~\cite{panero} which differ noticeably from the old
data~\cite{boyd}.

In this paper we investigate how the input information from
lattice QCD and the used forms of the effective potential
influence general properties of thermodynamics at finite
temperature $T$ and baryon chemical potential $\mu$. After
introduction, in Sect. 2 we consider the polynomial and
logarithmic parameterizations of the Polyakov loop effective
potential for the new and old pure gluon lattice data within the
two-flavor PNJL model. Independent of the temperature, the model
parameters defined by properties of quarks and mesons are
presented in Sect. 3. Comparative study of the thermodynamics and
phase structure, their dependence  on the lattice input and used
parametrization are considered  in Sect. 4 at finite $T$ and
$\mu$. The last Section summarizes the obtained results.

\subsection{The Nambu--Jona-Lasinio model with Polyakov-loop}

The deconfinement in the pure $SU(N_c)$ gauge theory can be
simulated by introducing an effective potential for a complex
Polyakov loop field. The PNJL
Lagrangian employed in this work is~\cite{ratti}
\begin{eqnarray}
\label{Lpnjl}
\mathcal{L}_{\rm PNJL}=\bar{q}\left(i\gamma_{\mu}D^{\mu}-\hat{m}_0
\right) q+ G \left[\left(\bar{q}q\right)^2+\left(\bar{q}i\gamma_5
\vec{\tau} q \right)^2\right]
-\mathcal{U}\left(\Phi[A],\bar\Phi[A];T\right)~.
\end{eqnarray}
Here,  a local chirally symmetric scalar-pseudoscalar four-point interaction of
quark fields $q,\bar{q}$ is introduced with an effective coupling strength $G$,
$\vec{\tau}$ is the vector of Pauli matrices in flavor space,
$\hat{m}_0$ is the diagonal matrix of the 2-flavor current quark masses,
$\hat{m}_0 = \mbox{diag} \, (m^0_u, m^0_d)$, $m^0_u = m^0_d=m_0$.

The quark fields are coupled to the gauge field $A^\mu$ through the covariant
derivative $D^\mu = \partial^\mu -iA^\mu$. The gauge coupling $g$ is conveniently
absorbed in the definition $A^\mu(x)=g{\cal A}_a^\mu\frac{\lambda_a}{2}$ where
${\cal A}_a^\mu$ is the $SU(3)$ gauge field and $\lambda_a$ is the Gell-Mann
matrices. The gauge field is taken in the Polyakov gauge
$A^\mu = \delta_0^\mu A^0 = -i\delta_4^\mu A_4$.
The field $\Phi$ is determined by the trace of the Polyakov loop $L(\vec{x})$ and
 its  conjugate~\cite{ratti}
\begin{eqnarray}
\Phi[A] = \dfrac{1}{N_c} \mbox{Tr}_c L(\vec{x})~,\hspace*{5mm}
\bar{\Phi}[A] = \dfrac{1}{N_c} \mbox{Tr}_c L^\dagger(\vec{x})~,\nonumber
\end{eqnarray}
where $L(\vec{x}) = \mathcal{P} \exp \left[ \displaystyle i
\int_{0}^{\beta} d \tau A_4 (\vec{x}, \tau) \right]$,
$\beta=1/T$ being the inverse temperature. In the absence of quarks, we
have $\Phi=\bar{\Phi}$ and the Polyakov loop servers as an order
parameter for deconfinement.

The gauge sector of the Lagrangian density (\ref{Lpnjl}) is
described by an effective potential
$\mathcal{U}\left(\Phi[A],\bar\Phi[A];T\right)$. The effective
potential must satisfy the $Z(3)$ center symmetry. In accordance
with the underlying $Z(3)$ symmetry, one can choose the following
general polynomial form:
\begin{eqnarray}\label{effpot}
\frac{\mathcal{U}\left(\Phi,\bar\Phi;T\right)}{T^4}
&=&-\frac{b_2\left(T\right)}{2}\bar\Phi \Phi-
\frac{b_3}{6}\left(\Phi^3+ {\bar\Phi}^3\right)+
\frac{b_4}{4}\left(\bar\Phi \Phi\right)^2, \\ \label{Ueff}
b_2\left(T\right)&=&a_0+a_1\left(\frac{T_0}{T}\right)+a_2\left(\frac{T_0}{T}
\right)^2+a_3\left(\frac{T_0}{T}\right)^3~.
\end{eqnarray}
The $Z(3)$ symmetry leads to some freedom in the choice of the
effective potential form. Along with the simplest polynomial form,
Eq.(\ref{effpot}), there exists an expression  with a logarithm in
place of the higher order polynomial terms in $\bar{\Phi}$, $\Phi$
\cite{Fu04}.  In the logarithmic form the potential is
\begin{eqnarray}\label{effpot_log}
\frac{\mathcal{U}\left(\Phi,\bar\Phi;T\right)}{T^4}
&=&-\frac{1}{2}a\left(T\right)\bar\Phi \Phi+
b\left(T\right){\rm ln}\left[1-6\bar{\Phi}\Phi + 4(\bar{\Phi}^3+\Phi ^3) - 3(\bar{\Phi}\Phi)^2\right], \\ \label{Ueff_log}
a\left(T\right)&=&\tilde{a}_0+\tilde{a}_1\left(\frac{T_0}{T}\right)+
\tilde{a}_2\left(\frac{T_0}{T}
\right)^2, b\left(T\right) = \tilde{b}_3\left(\frac{T_0}{T}\right)^3.
\end{eqnarray}
The pressure of a pure-gauge system is given by
$p=-\mathcal{U}$.

Pure-gluon lattice data from Ref.~\cite{boyd} are traditionally
used to find the parameter set for both forms of the effective
potential ~\cite{ratti,ratti_log}. In contrast, our work is based
on new gluon lattice data~\cite{panero} looking for a new potential
to fit the lattice pressure. In finding the potential
parameters the following conditions should be satisfied: $\Phi
\rightarrow 1$  and $p/T^4\rightarrow 1.75, $  when
$T\rightarrow\infty$. As immediately follows from these
conditions,  $\tilde{a}_0 = 3.51$ for the logarithmic potential
and the constraint $1.75 = a_0/2 + b_3/3 - b_4/4$ for the
polynomial potential.  Minimizing $\mathcal{U}(\Phi, \bar{\Phi},
T)$ with respect to variation of $\Phi$ and taking into account 
that $\Phi = \bar{\Phi}$ at $\mu= 0$,  
we can find parameters using the  method of least mean
squared deviations.  Thus, for the critical temperature $T_0=$270 MeV
the following parameter sets were obtained (see Table \ref{table1}
and Table \ref{table11}).

\begin{table}[bh]
\tbl{  Parameters of the effective potential  $\mathcal{U}[A]$ with the
polynomial form.}
{\begin{tabular}{@{}ccccccc@{}} \toprule
&$a_0$ & $a_1$ & $a_2$ & $a_3$ & $b_3$ & $b_4$ \\ \colrule
old data~\cite{boyd} &6.75 & -1.95 & 2.625 & -7.44 & 0.75 & 7.5 \\ \botrule
new data~\cite{panero} &6.47 & -4.62 & 7.95 & -9.09 & 1.03 & 7.32 \\ \botrule
\end{tabular}\label{table1}}
\end{table}
\begin{table}[bh]
\tbl{  Parameters of the effective potential
$\mathcal{U}\left(\Phi,\bar\Phi;T\right)$ in the logarithmic form.}
{\begin{tabular}{@{}ccccc@{}} \toprule
&$\tilde{a}_0$ & $\tilde{a}_1$ & $\tilde{a}_2$ & $\tilde{b}_3$ \\ \colrule
old data~\cite{boyd} &3.51 & -2.47 & 15.2 & -1.75   \\ \botrule
new data~\cite{panero} &3.51 & -5.121 & 20.99 & -2.09 \\ \botrule
\end{tabular}\label{table11}}
\end{table}

\begin{figure}[b]
\centerline{
\includegraphics[width = 11.5cm]{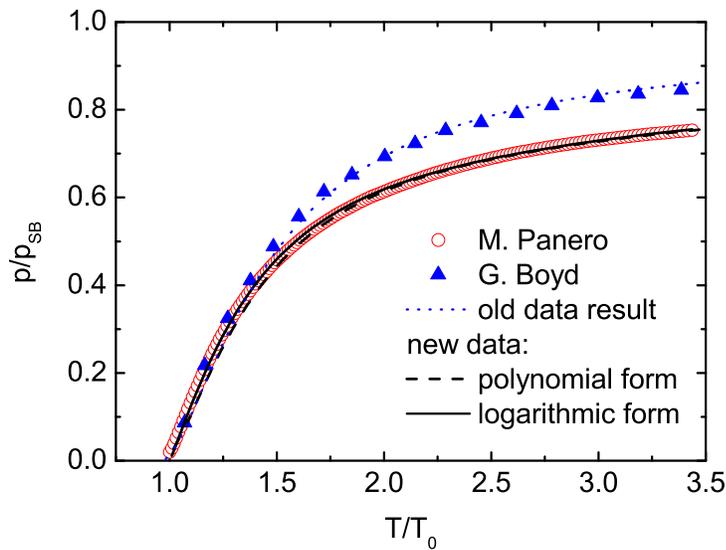}
}
\caption{Scaled pressure in the pure gauge sector as function of scaled temperature.
The old$^{22}$ 
and new$^{21}$
lattice data are plotted by circles and
triangles, respectively. Solid lines correspond to polynomial form of potential and
dashed lines correspond to the logarithm form. }
\label{pres_lat}
\end{figure}

In Fig. \ref{pres_lat},  old and new lattice gluon data are
compared together with the results of their approximations. As is
seen the lattice results differ by about 10$\%$ at $T/T_0\gsim 2$.
New data are plotted by circles but the density of measured points
is so high that the results look like a shaded band. Both
polynomial and logarithmic forms are in nice agreement with the
data and it is hard to distinguish them from each other.
\begin{figure}[h]
\centerline{
\includegraphics[width = 7.0cm]{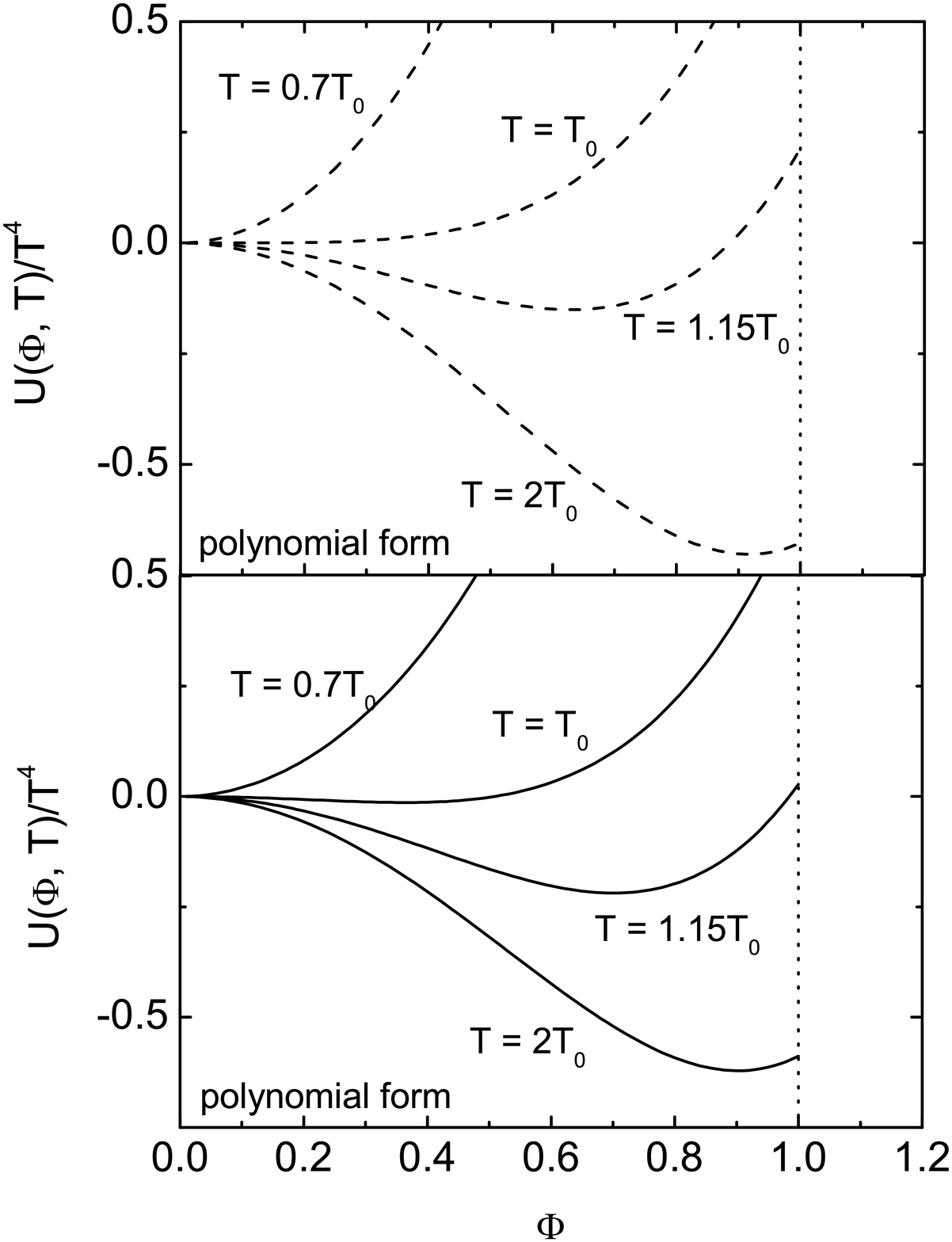}
\includegraphics[width = 7.0cm]{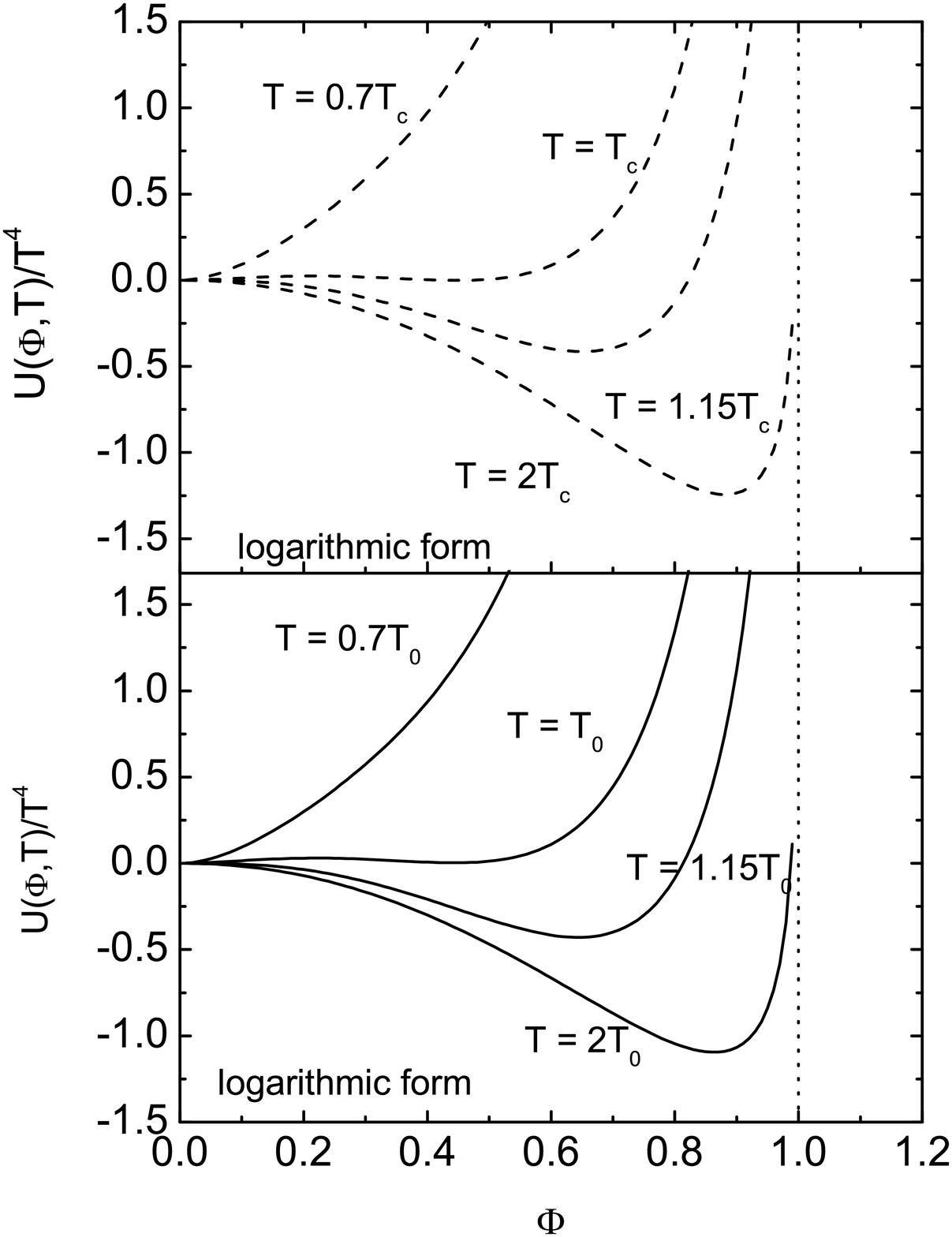}\hspace*{-10mm}
}
\caption{The Polyakov loop effective potential $\mathcal{U}$ as a function of
$\Phi$ for various values of temperature for old (top) and new (bottom) sets of
parameters. Left panel corresponds to the polynomial form  and right
panel corresponds to the logarithm form of the potential.}
\label{ef_pot}
\end{figure}

In general, the parameter $T_0$ depends on the number of active
flavors and the chemical potential \cite{Schaefer:2007pw}.  In the
pure gauge sector $T_0 = 0.27$ GeV was used \cite{panero}. The
effective potential for both sets of parameters at $T = 0.2$,
$0.27$, $0.32$, $0.54$ GeV is shown in Fig. \ref{ef_pot}. Both 
sets  describe quite satisfactorily the Polyakov loop as a
function of temperature. In accordance with the $Z(3)$ center
symmetry, the following properties of the effective potential
$\mathcal{U}(\Phi,\bar{\Phi}; T)$ are seen. At low temperature
$\mathcal{U}(\Phi,\bar{\Phi}; T)$ has a single minimum at $\Phi =
0$ (a confinement phase); the effective potential is getting flat
for the critical temperature $T=T_0$ and above critical
temperature (a deconfinement phase) a second minimum arises at nonzero
$\Phi$, as a consequence of $Z(3)$ symmetry breaking; in the
$T\rightarrow \infty$ limit, $\Phi \rightarrow 1$ (see Fig.
\ref{ef_pot}). One should note that after the second minimum  the
logarithmic potential $\Phi$ increases faster than the polynomial
one forming a more distinct minimum. With the introduction of
quarks the critical temperature goes down. The range of
applicability of this model is $T\lsim 2.5 T_c$ since at higher
temperature transverse gluons start to contribute
significantly~\cite{MOM04}.

\begin{figure}[h]
\centerline{
\includegraphics[width = 7.cm]{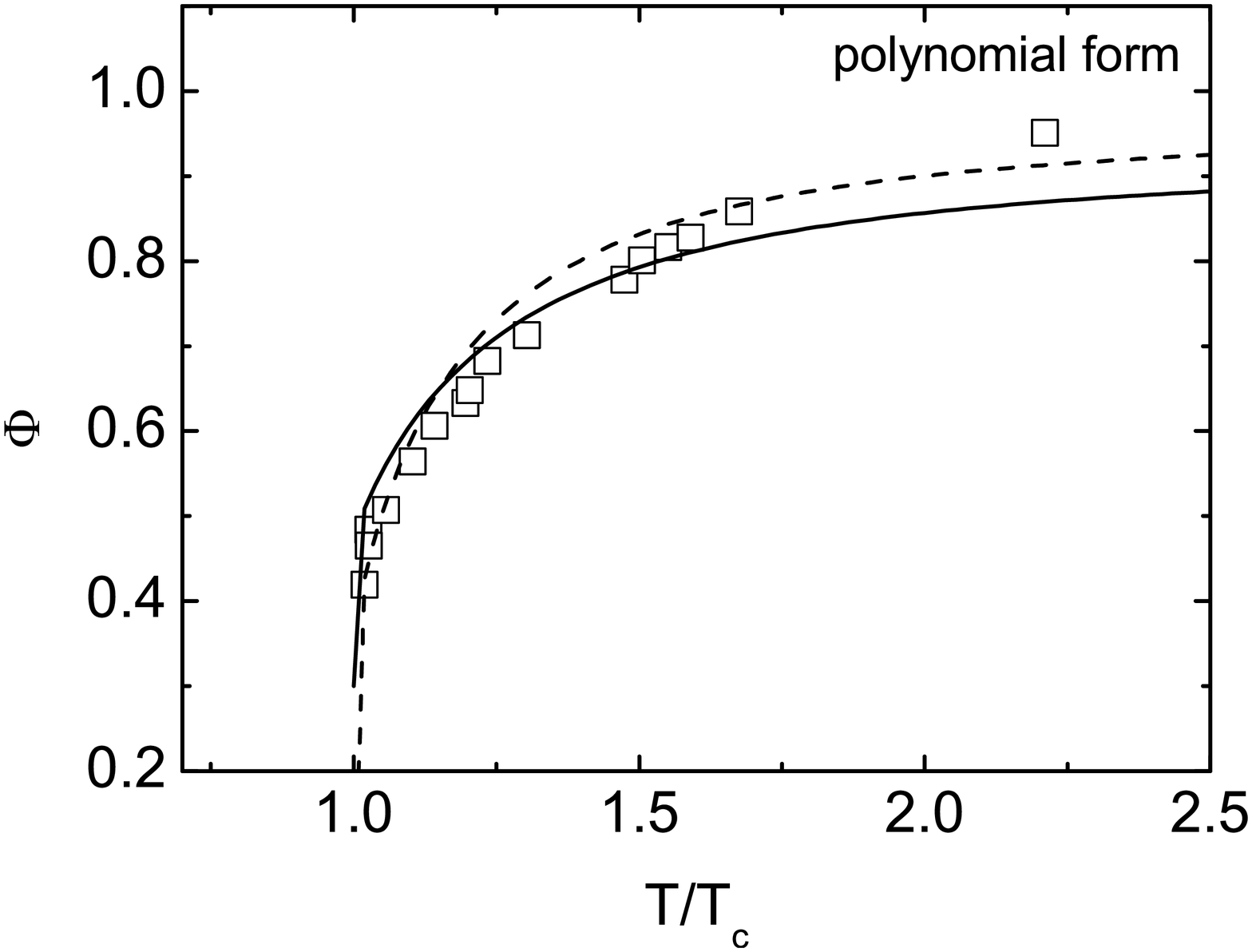}\hspace*{-10mm}
\includegraphics[width = 7.cm]{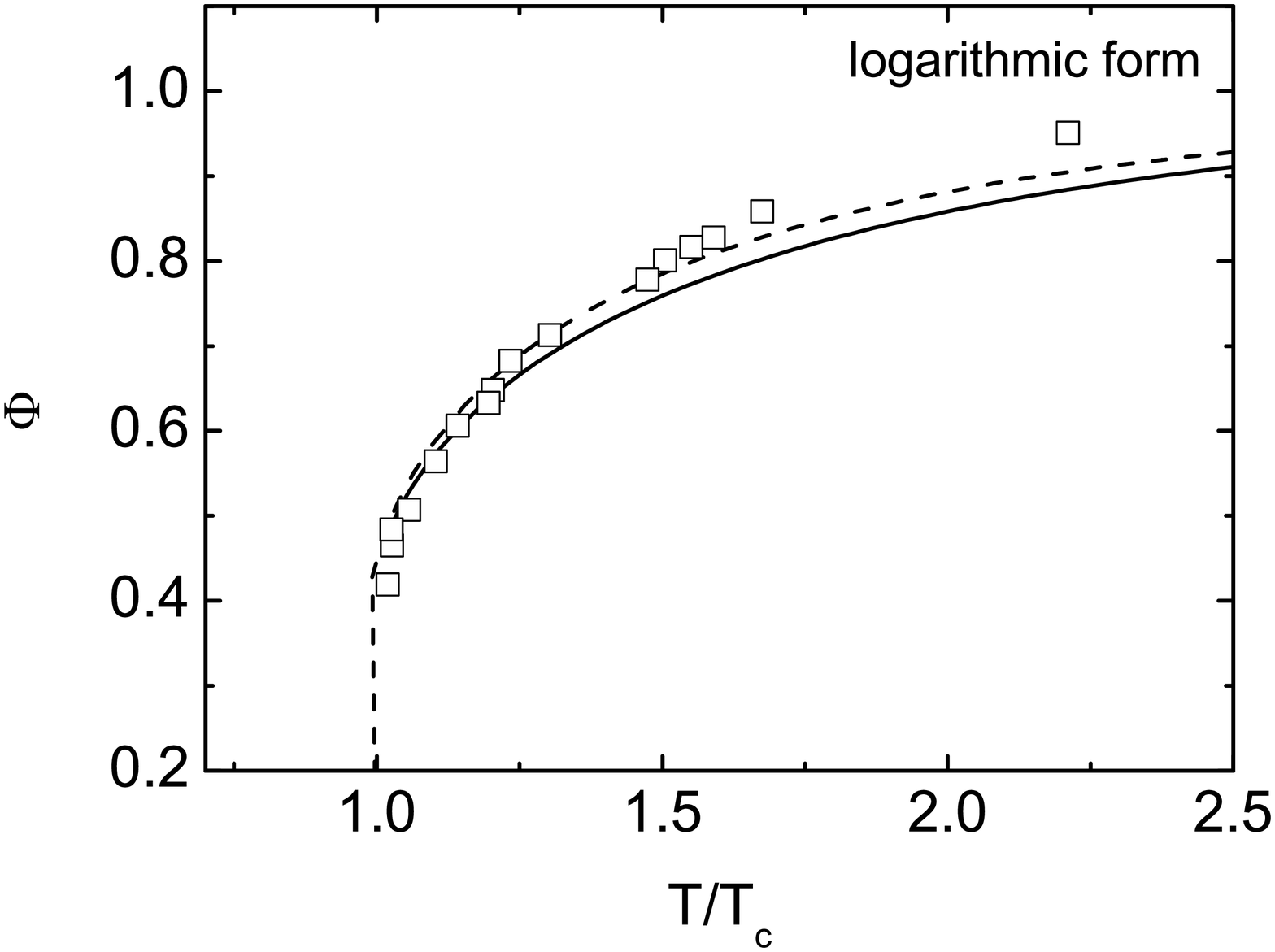}
}
\caption{Temperature dependence of the Polyakov loop $\Phi$ for polynomial
(\ref{effpot}) (left) and logarithmic  (\ref{effpot_log}) (right) forms. Lattice
data are from $^{25}$
Solid and dashed lines correspond to new and old parameter sets, respectively.
}
\label{Ploop}
\end{figure}

The Polyakov loop is compared with the lattice results  in
Fig.~\ref{Ploop}. In reasonable agreement of both forms with the
lattice data, the new parameter set predicts slightly lower values
of the field $\Phi$ because the pressure for new data is also
below the old one.

\section{Quarks and light mesons in the PNJL model}

The grand  potential density for the PNJL ($N_f=$2) model in the
mean-field approximation is given by the following
equation~\cite{ratti,hansen}:

\begin{eqnarray}
\label{potpnjl}
\Omega (\Phi, \bar{\Phi}, m, T, \mu) &=&
\mathcal{U}\left(\Phi,\bar\Phi;T\right) + G \langle \bar{q}q \rangle ^2
+\Omega_q~,
\end{eqnarray}
where the quark term is
\begin{eqnarray}
\Omega_q = -2 N_c N_f \int \dfrac{d^3p}{(2\pi)^3} E_p
- 2N_f T \int \dfrac{d^3p}{(2\pi)^3} \left[ \ln N_\Phi^+(E_p)+
\ln N_\Phi^-(E_p) \right]~
\label{omegaq}
\end{eqnarray}
 and the functions are
\begin{eqnarray}
&& N^+_\Phi(E_p) = \left[ 1+3\left( \Phi +\bar{\Phi} e^{-\beta
E_p^+}\right) e^{-\beta E_p^+} + e^{-3\beta E_p^+}
\right], \\
&& N^-_\Phi(E_p) = \left[ 1+3\left( \bar{\Phi} + {\Phi} e^{-\beta
E_p^-}\right) e^{-\beta E_p^-} + e^{-3\beta E_p^-} \right]~,
\end{eqnarray}
where $E_p=\sqrt{{\bf p}^2+m^2}$ is the quasiparticle energy of
the quark; $E_p^\pm = E_p\mp \mu$, the upper sign applying for
fermions and the lower sign for antiparticles.

Since NJL-type models are nonrenormalizable, it is necessary to
introduce a regularization, e.g., by a cutoff $\Lambda$ in the
momentum integration. Following \cite{hansen}, we use in this
study the three-dimensional momentum cutoff $\Lambda$ for vacuum
terms  and extend this integration to infinity for the matter 
contributions given by the second term of Eq. \ref{omegaq}. 
A comprehensive study of the differences between the
two regularization procedures (with and without cutoff on the
quark momentum states at finite temperature) was performed in
\cite{costa2}.

In the mean-field approximation, we can obtain the constituent
quark mass $m$ from the condition that the thermodynamic potential
(\ref{potpnjl})  will have a minimum  with respect to variation of
this parameter, $\partial \Omega/\partial m = 0$. This condition
is equivalent to the gap equation~\cite{hansen,klevansky}
\begin{eqnarray}
\label{gap1}
m = m_0 -2 G \ \langle \bar{q} q \rangle \label{masq}~,
\end{eqnarray}
where the quark condensate is defined as
$\langle \bar{q} q \rangle = \partial \Omega/\partial m_0$.
For the mass gap equation we get
\begin{eqnarray}
m = m_0 + 4 G N_c N_f \int_{\Lambda} \dfrac{d^3p}{(2\pi)^3}
\dfrac{m}{E_p} \left[ 1 - f^+ - f^- \right]
\end{eqnarray}
with the modified Fermi-Dirac distribution functions for fermions and antifermions
\begin{eqnarray}
f^+ &=& \left[\left( \Phi +2\bar{\Phi} e^{-\beta E_p^+}\right) e^{-\beta E_p^+}
+ e^{-3\beta E_p^+}\right]/ N_\Phi^+(E_p)~,
\label{fermimod}\\
f^- &=& \left[\left(\bar{\Phi}+2{\Phi} e^{-\beta E_p^-}\right) e^{-\beta E_p^-}
+ e^{-3\beta E_p^-} \right]/ N_\Phi^-(E_p)~.
\label{afermimod}
\end{eqnarray}
Moreover, for PNJL calculations we should find the values of
$\Phi$ and $\overline{\Phi}$ by minimizing $\Omega$ with respect
to $\Phi$ and $\overline{\Phi}$~\cite{hansen} at given $T$ and
$\mu$. One should note that if $\Phi \rightarrow 1$, the
expressions Eqs.~(\ref{fermimod}),(\ref{afermimod}) reduce to the
standard NJL model.

For a self-consistent description of the particle spectrum in the mean-field
approximation, the meson correlations have to be taken into consideration.
These correlations are related to the polarization operator of constituent
fields.
For scalar and pseudoscalar particles the polarization operators are
represented by loop integrals~\cite{schulze,quack,eqpi}

\begin{eqnarray}
\Pi^{PP}_{ab} (P^2) &=& \int \frac{d^4p}{(2\pi)^4} \mbox{Tr}\,
\left[ i \gamma_5 \tau^a S(p+P) i \gamma_5 \tau^b S(p)
 \right], \label{Polpi} \\
\Pi^{SS}_{ab} (P^2) &=& \int \frac{d^4p}{(2\pi)^4} \mbox{Tr}\,
 \left[S(p+P) S(p) \right],
  \label{Polsig}
\end{eqnarray}
where $S(p)$ is the quark propagator and the operation $\mbox{Tr}$ is taken 
over Dirac, flavor and color indices of quark fields.

From the point of view of the polarization operators, the
pseudoscalar ($\pi$) and scalar ($\sigma$) meson masses can be
defined by
 the condition that for $P^2=M_\pi^2$ $(M_\sigma^2)$ the corresponding
polarization operator $\Pi^{PP}(M_\pi^2)$ $(\Pi^{SS}(M_\sigma^2))$ leads to a
bound state pole in the corresponding meson correlation function \cite{hansen}.
For mesons at rest (${\mathbf P}=0$) in the medium, these conditions
correspond to the equations
 \begin{eqnarray}
1 + 16 G N_c N_f \int \frac{d^3 p}{(2\pi)^3}
\frac{E_p}{M_\pi^2-4 E_p^2} \left( 1- f^+ - f^- \right) &=& 0,
\label{masspi}\\
1 + 16 G N_c N_f \int \frac{d^3 p}{(2\pi)^3}\frac{1}{E_p}
\frac{E_p^2-m^2}{M_\sigma^2-4 E_p^2} \left(1-f^+ -f^-\right) &=& 0~.
\label{masssigma}
\end{eqnarray}

\begin{figure}[h]
\centerline{
\psfig{file = 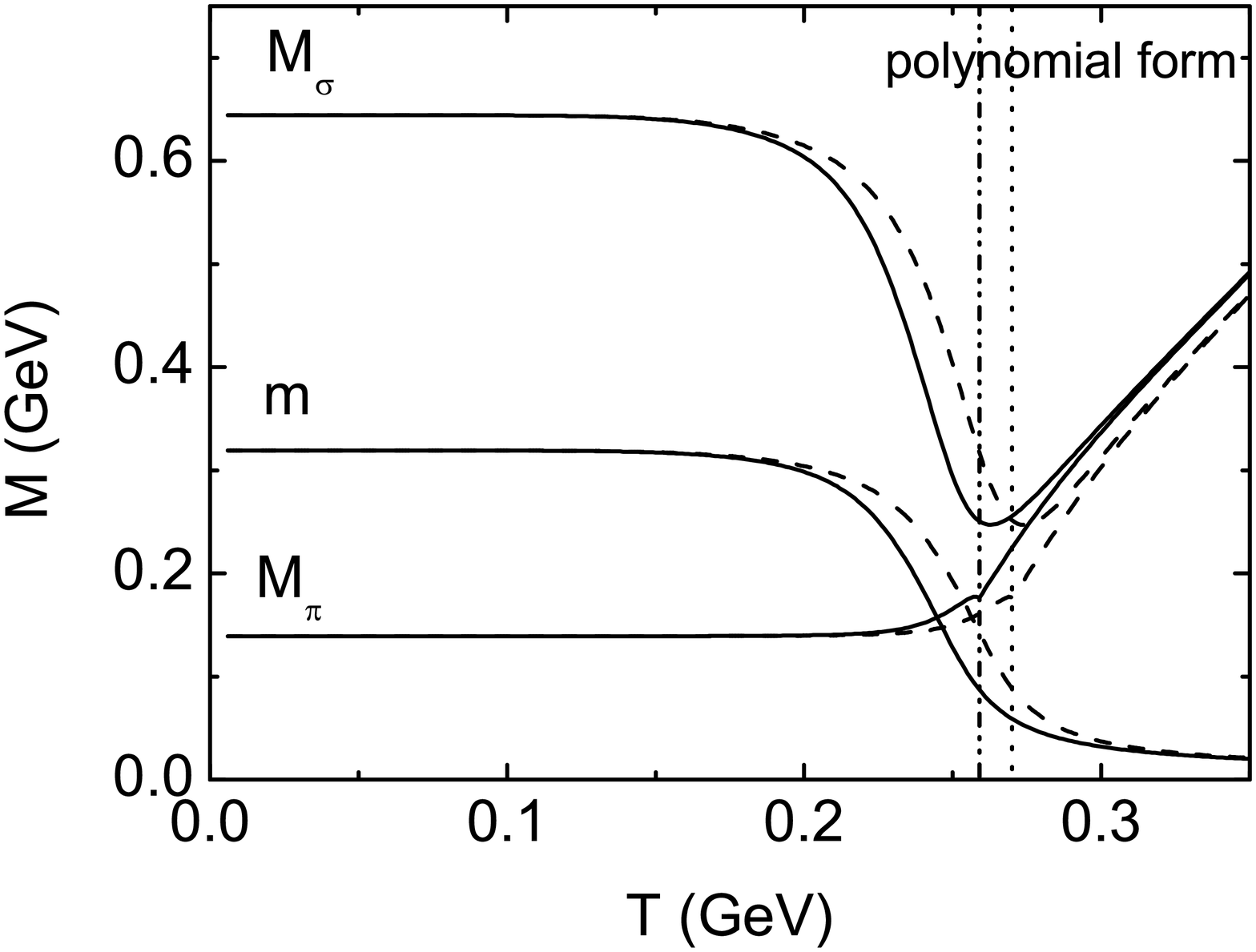, width = 7.5 cm}\hspace{-10mm}
\psfig{file = 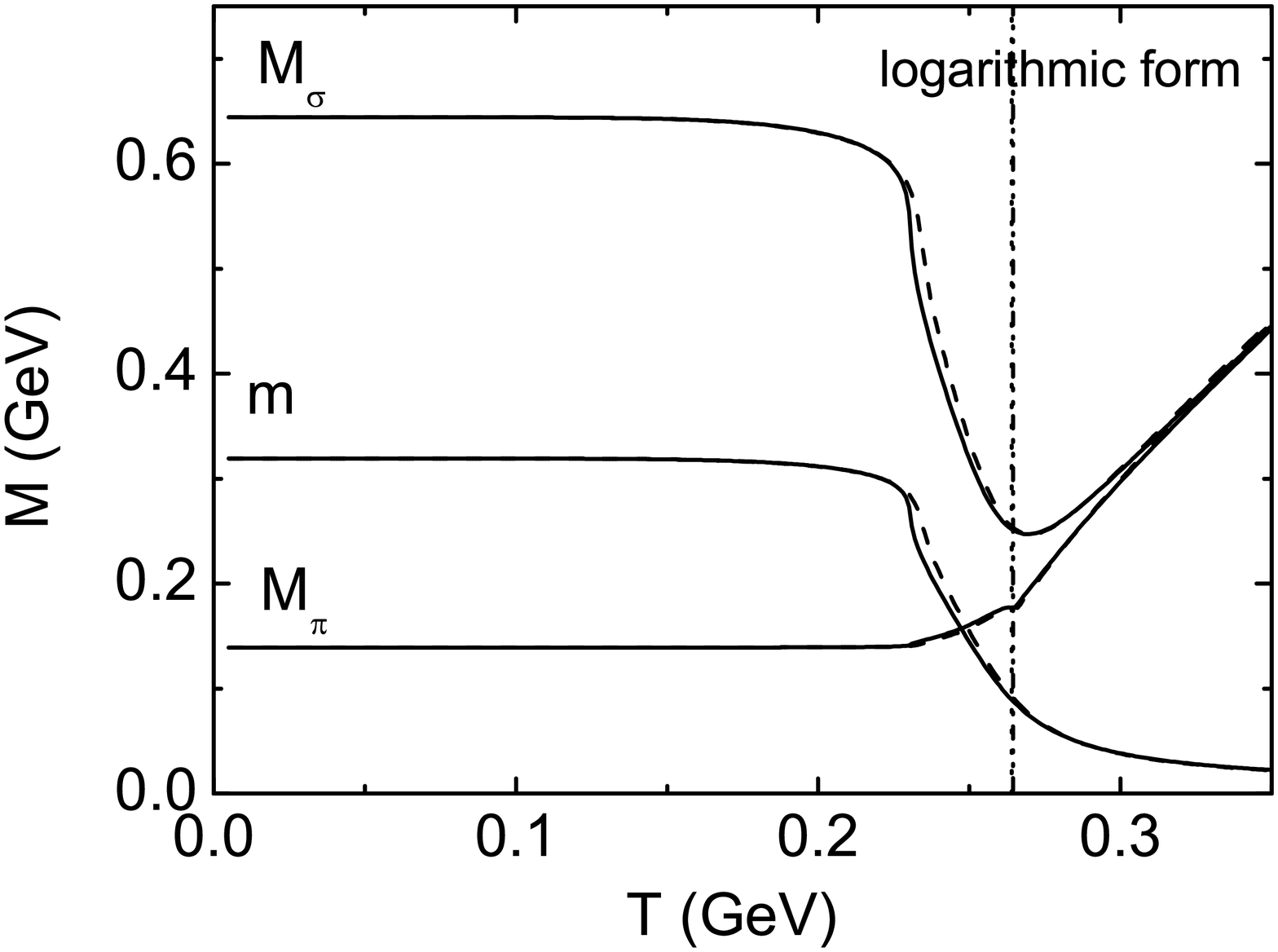, width = 7.5 cm}}
\caption{ Temperature dependence of the masses $m_q$, $M_\pi$ and $M_\sigma$ masses
at $\mu$ = 0 GeV for polynomial (left panel) and logarithmic (right panel) forms
of potential. The PNJL results for new and old parameter sets are given
by the solid and dashed lines, respectively. The Mott temperatures for both
parameter sets are plotted by vertical lines.}
 \label{masses}
\end{figure}

In order to solve Eqs.~(\ref{masq}), (\ref{masspi}) and
(\ref{masssigma}), a set of model parameters has to be determined:
the above-mentioned cutoff parameter $\Lambda$, the current quark
mass $m_0$ (in the chiral limit $m_0=0$) and the coupling constant
$G$. These parameters are fixed at $T = 0$ to reproduce physical
quantities: the pion mass $M_\pi = 0.139$ GeV, the pion decay
constant $F_\pi = 0.092$ GeV and the quark condensate $\langle
\bar{q} q\rangle^{1/3}=-250$ MeV. The obtained parameters are the
same as in those obtained earlier in our
papers~\cite{FKT11,FKT11-2} and are shown in Table~\ref{table2}.
\begin{table}[h]
\tbl{The set of model parameters reproducing observable quantities
(in brackets) and the chiral condensate
$\langle \bar{q} q\rangle^{1/3}=-250$ MeV.}
{\begin{tabular}{@{}ccccc@{}}\toprule
$m_0$ [MeV] & $\Lambda$ [GeV] & $G$ [GeV]$^{-2}$ & $F_\pi$ [GeV] & $M_\pi$ [GeV] \\ \colrule
5.5 & 0.639 & 5.227 & (0.092) & (0.139) \\
\botrule
\end{tabular}
 \label{table2}}
 \end{table}
Meson masses obtained as solutions of the gap-equation
(\ref{masq}) and Eqs.~(\ref{masspi}), (\ref{masssigma}) at nonzero
$T$ are presented in Fig.~\ref{masses}  for two parameter sets.

The temperature modification of the quasiparticle properties is
clearly seen in this figure. Up to the Mott temperature $T_{\rm Mott}$, 
defined as  $M_\pi (T_{\rm Mott})= 2m_q(T_{\rm
Mott})$, the $\sigma$ mass practically follows the behavior of
$2m_q(T)$ with a drop toward the pion mass signaling partial chiral
symmetry restoration. At $T > T_{\rm Mott}$ the masses of chiral
partners become equal to each other, $M_\sigma \approx M_\pi$, and
then both masses monotonically increase with temperature.  Below
the Mott temperature, the pion mass remains practically constant.
It justifies that $T_{\rm Mott}$ is a little bit lower for the
new parameterization than for the old one when the polynomial form is
used, while both values of $T_{\rm Mott}$ coincide in the case of
logarithmic parametrization.

\section{Thermodynamics of the PNJL models }

The thermodynamics of particles is described in terms of the grand canonical
ensemble which is related to the Hamiltonian $H$ as follows:
\begin{eqnarray}
 \label{can}
e^{-\beta V \Omega} = \mbox{Tr}\,\, e^{-\beta (H-\mu N)},
\end{eqnarray}
where $N$ is the particle number operator, $\mu$ is the quark chemical
potential and the operator $\mbox{Tr}$ is taken over momenta
as well as color, flavor and Dirac indices. If $\Omega$ is known,
the basic thermodynamic quantities - the pressure $p$, the energy density
$\varepsilon$, the entropy density $s$, the density of quark number
$n$ and the specific heat  $c_v$ - can be defined as follows:
\begin{eqnarray} \label{p}
p &=& -\frac{\Omega}{V}, \\ \label{s}
s &=& -\left(\frac{\partial \Omega}{\partial T}\right)_\mu, \\ \label{epsilon}
\varepsilon &=& -p + Ts +\mu \, n, \\ \label{n}
n &=& -\left(\frac{\partial \Omega}{\partial \mu}\right)_T, \\ \label{Cv}
c_v &=& \frac{T}{V}\left(\frac{\partial s}{\partial T}\right)_\mu ~.
\end{eqnarray}
\begin{figure}[h]
\centerline{
\psfig{file = 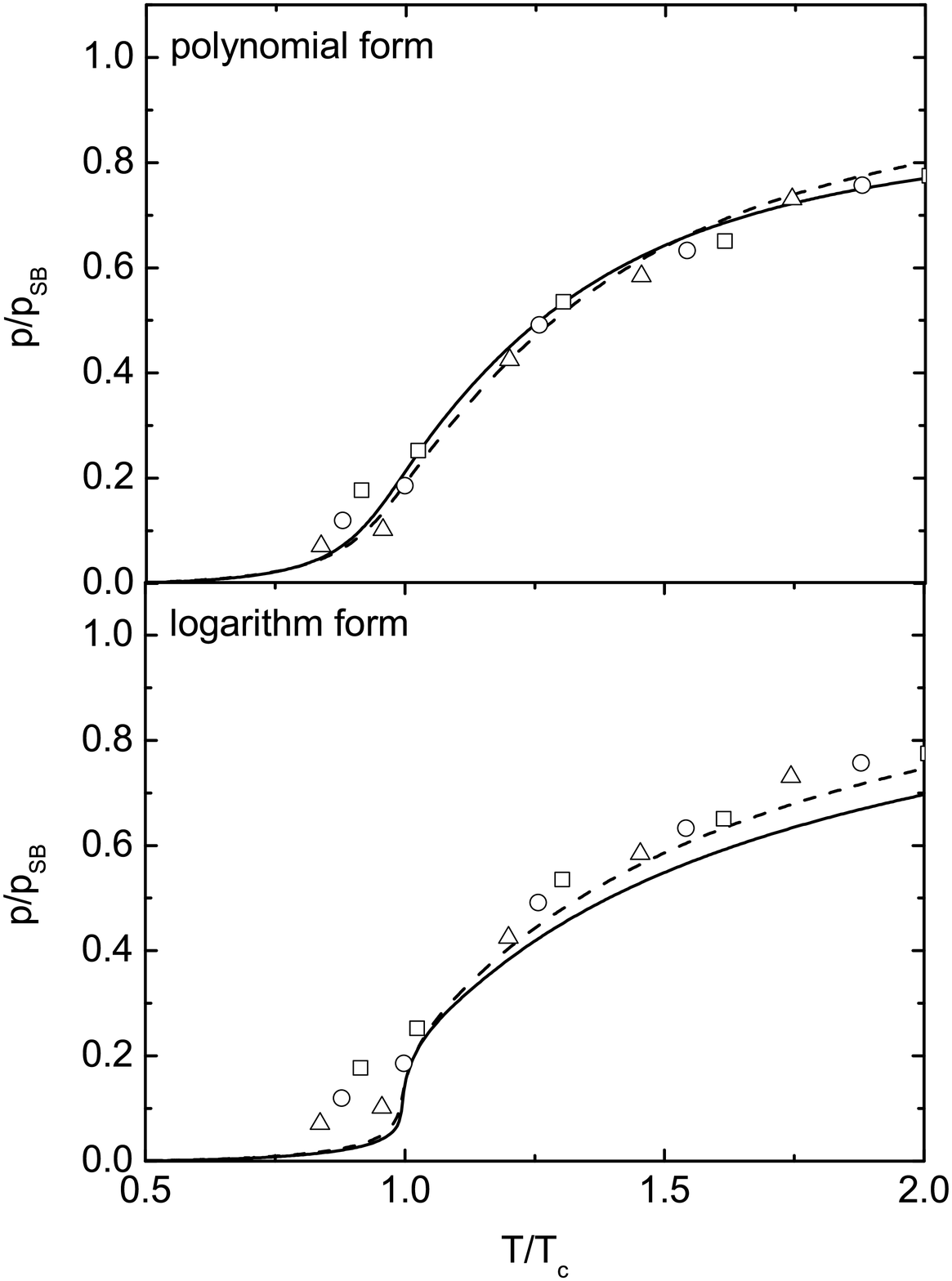, width=6.5cm}\hspace*{1mm}
\psfig{file = 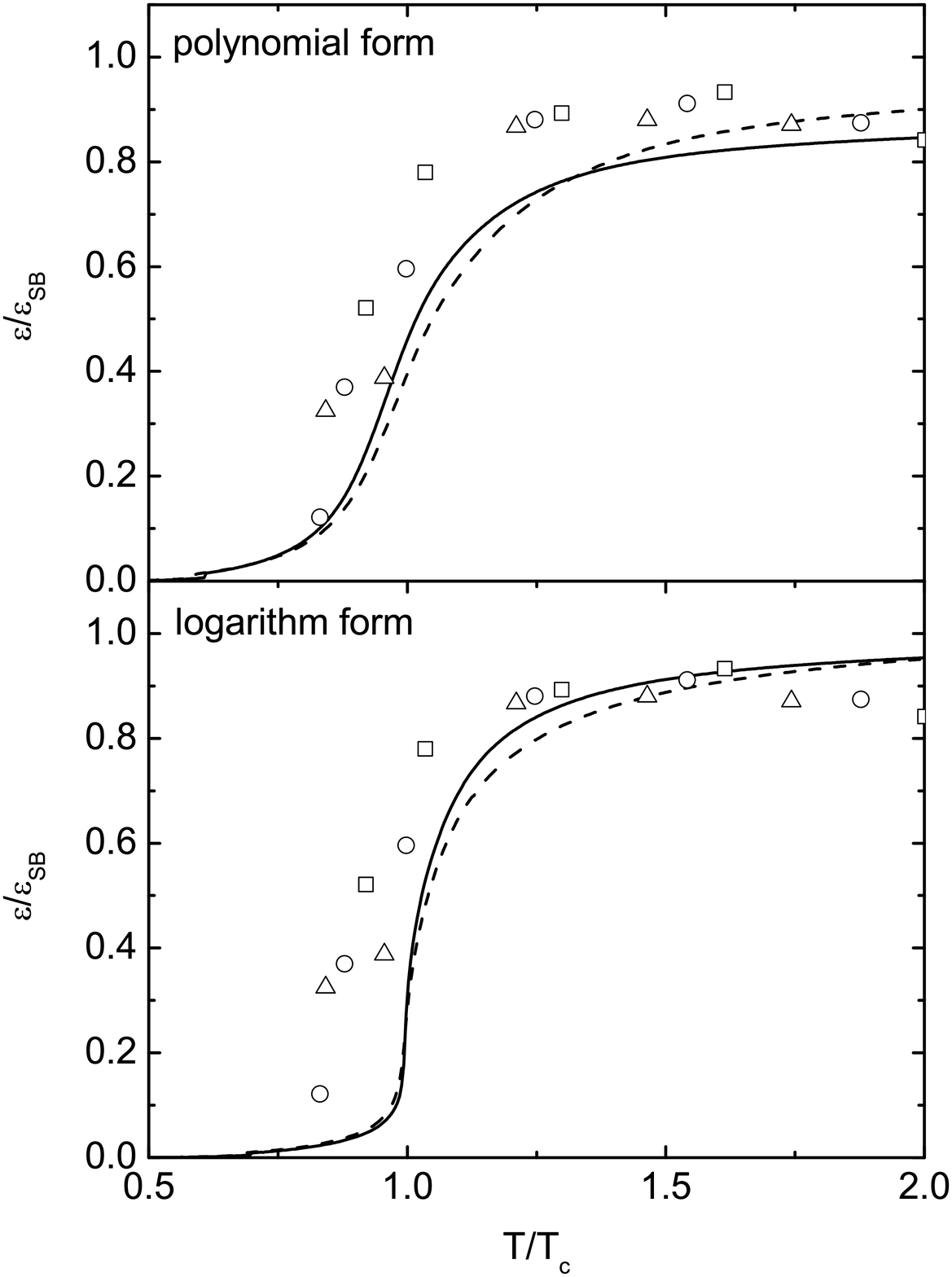, width=6.5cm} }
\caption{The temperature dependence of the  scaled pressure
and energy density within the PNJL model for $\mu=0$
in two schemes of parametrization.  Dotted lines correspond to the old data
parameterization and solid lines are new ones.
Lattice data points for $N_f=2$ at $\mu=$0 are
from Ref. 30.
Circles, squares and diamonds correspond to calculations at $N_t=$6
with the mass ratio of the pseudoscalar to vector meson
$m_{PS}/m_{V}=$0.65, 0.70 and 0.75, respectively.
}
\label{therm}
\end{figure}
The thermodynamic potential in equilibrium corresponds to a global minimum
with respect to variations of the order parameter
\begin{eqnarray}
\frac{\partial \Omega (T,\mu,m) }{\partial m} = 0, \,\,\,
\frac{\partial^2 \Omega(T,\mu,m)}{\partial m^2} \geqslant 0.
\end{eqnarray}

All these relations (\ref{p}),(\ref{s}),(\ref{epsilon}),(\ref{n}),(\ref{Cv})
describe the thermodynamics of the system. For the
considered models the thermodynamic potentials are defined from
Eq. (\ref{potpnjl}). From this equation we can read off the vacuum
part

\begin{eqnarray}
\label{omvac}
\Omega_{vac} = \frac{(m -m_{0})^2}{4G} - 2N_c N_f\int
\frac{d^3p}{(2\pi)^3}E_p.
\end{eqnarray}
This quantity does not vanish as $T\rightarrow 0$ and $\mu
\rightarrow 0$. Therefore, in order to obtain the physical
thermodynamical potential which corresponds to vanishing pressure
and energy density at $(T,\mu)=(0,0)$, one has to renormalize the
thermodynamic potential by subtracting its vacuum expression
(\ref{omvac}). This corresponds to the following definition of the
physical pressure
\begin{eqnarray}
\frac{p}{T^4} = \frac{p(T, \mu, m) - p(0, 0, m)}{T^4}.
\end{eqnarray}

With increasing temperature the pressure has to reach the
Stefan-Boltzmann limit~\cite{costa2} which in the chiral limit for
the PNJL model is given as
\begin{eqnarray}
\frac{p_{SB}}{T^4} = (N_c^2 - 1)\frac{\pi^2}{45} + N_cN_f\frac{7\pi^2}{180}
\simeq 4.053~,
\end{eqnarray}
where the first and second terms correspond to gluons and quarks, respectively.

If the regularization $\Lambda =$ 0.639 is used, the $T$-behaviour of the
thermodynamic quantities considered is roughly the same while their absolute
values are noticeably lower, being far from the Stefan-Boltzmann
limit~\cite{costa2}.
\begin{figure}[thb]
\centerline{
\psfig{file = 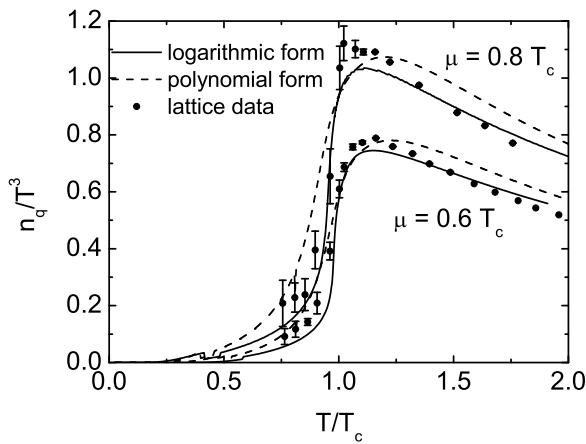, width=8.cm} }
\caption{Comparison of the scaled quark number density  as a function
of temperature at $\mu=$0.8 and 0.6 GeV. Lattice data  points are from
Ref. 31.
}
\label{nq}
\end{figure}

Within the PNJL model with $\Lambda\rightarrow\infty$ (we can use
$\Lambda\rightarrow\infty$ because most of the integrals in the
PNJL are convergent) the reduced pressure and energy density
exhibit reasonable behavior consistent with the recent lattice QCD
results for the vanishing chemical potential~\cite{AKh01} (see
Fig.~\ref{therm})  keeping in mind that the $m_{PS}/m_{V}$ ratio
in lattice calculations is still far from that for physical masses
$m_{PS}/m_{V}\sim 0.2$. As another example of thermodynamic
characteristics, the temperature dependence of the reduced quark
number density $n_q/T^3$ (see Eq.(\ref{n})) is presented in Fig.
\ref{nq}. Model results are in good agreement with the lattice
data for both values of the chemical potential considered. The
logarithmic approximation of the effective potential $\mathcal{U}$
seems to describe lattice data better than the polynomial one.

\subsection{Phase diagram and the CEP}

 Within NJL-like models there are several characteristic temperatures.
The parameter $T_0$ entering into the effective potential
(\ref{Ueff}) of the PNJL model has been noted above. For
$\pi$-mesons, the Mott temperature $T_{\rm Mott}$ is provided by
the condition $M_\pi(T_{\rm Mott}) = 2m_q(T_{\rm Mott})$. Above
$T_{\rm Mott}$ the pion dissociates into a quark and antiquark and
does not exist as a bound state. Similarly the $\sigma$ meson
dissociation temperature $T_d^\sigma$ is given by the equation
$M_\sigma(T_d^\sigma) = 2M_\pi(T_d^\sigma)$~\cite{quack,fu}.
 Other characteristics of phase transitions are the pseudo-critical temperature
for the chiral crossover $T_\chi$, defined by the maximum of
$\partial \langle q\overline{q} \rangle /\partial T$, and the
pseudo-critical temperature for the  crossover deconfinement
transition $T_p$ that can be found from the maximum of $\partial
\overline{\Phi}/\partial T$. Their difference is less than 0.013
but it increases with decreasing $T_0$.  The  third quantity is
$T_c$ assumed to equal  $T_\chi$. But in the PNJL model it is
higher than the lattice result $T_c \sim$ 192 GeV. Thus, it was
suggested to define $T_c$ as an average of two transition
temperatures $T_\chi$ and $T_p$~\cite{ratti,rosner}.

All these quantities obtained at $\mu=0$ are presented in Table \ref{table3}.
\begin{table}[h]
\tbl{Characteristic temperatures in the PNJL models for $\mu=$0.}
{\begin{tabular}{@{}ccccccc@{}} \toprule
 &  & $T_\chi$ &$T_p $&$ T_c $&$T_{\rm Mott}$ & $T^\sigma_d$ \\\colrule
 Polynomial form of potential & new & 0.2455 & 0.2335 & 0.2395 &
0.259 & 0.247\\
  & old & 0.2575 & 0.2485 & 0.253 & 0.27 & 0.257\\
 Logarithmic form of potential & new & 0.2305 & 0.2295 & 0.23
& 0.264 & 0.2645 \\
  & old & 0.2345 & 0.2335 & 0.234 & 0.2645 & 0.252 \\ \botrule
\end{tabular}
\label{table3}}
\end{table}

\begin{figure}[thh]
\centerline{ \psfig{file = 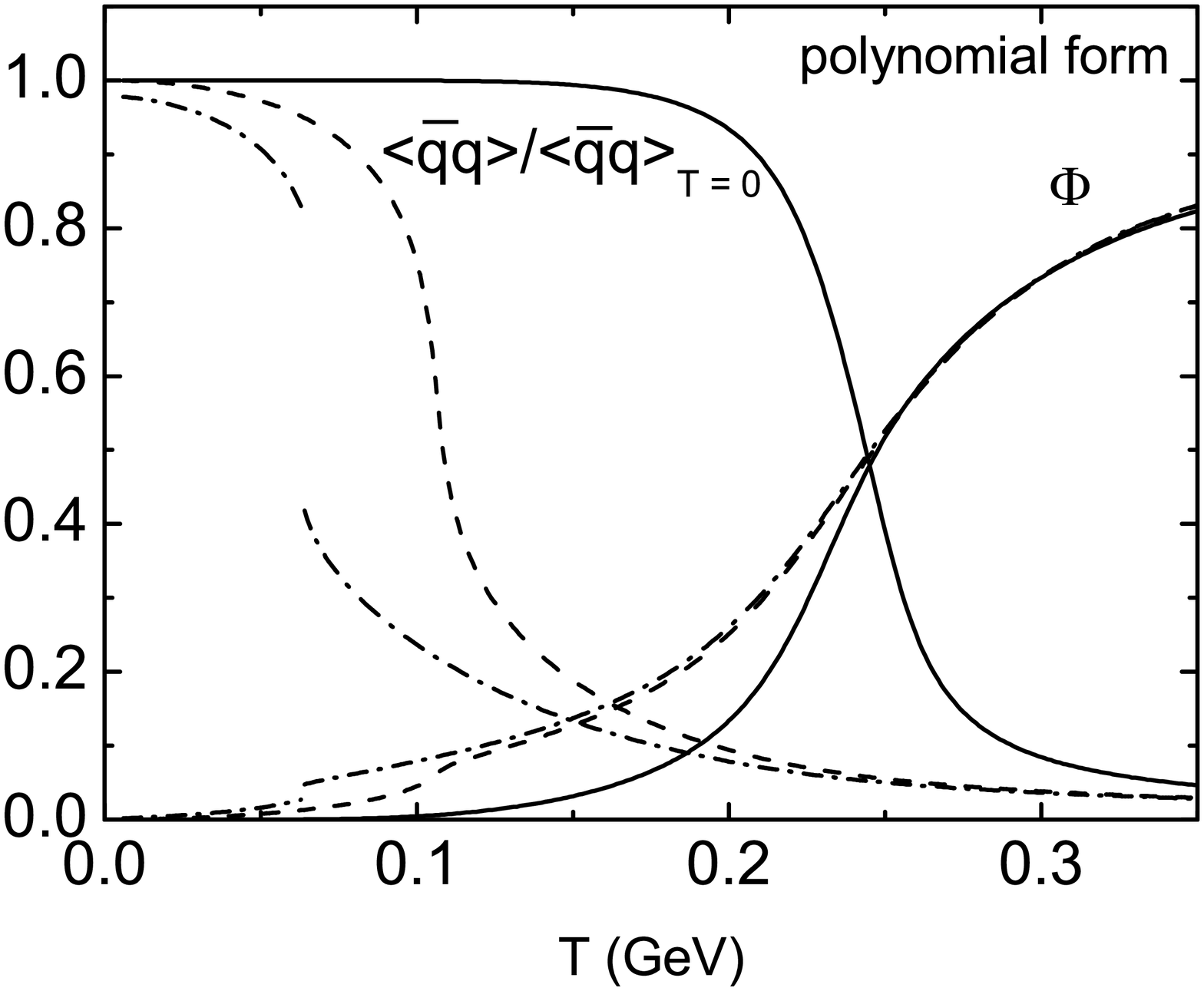, width =
7.cm}\hspace*{-10mm} \psfig{file = 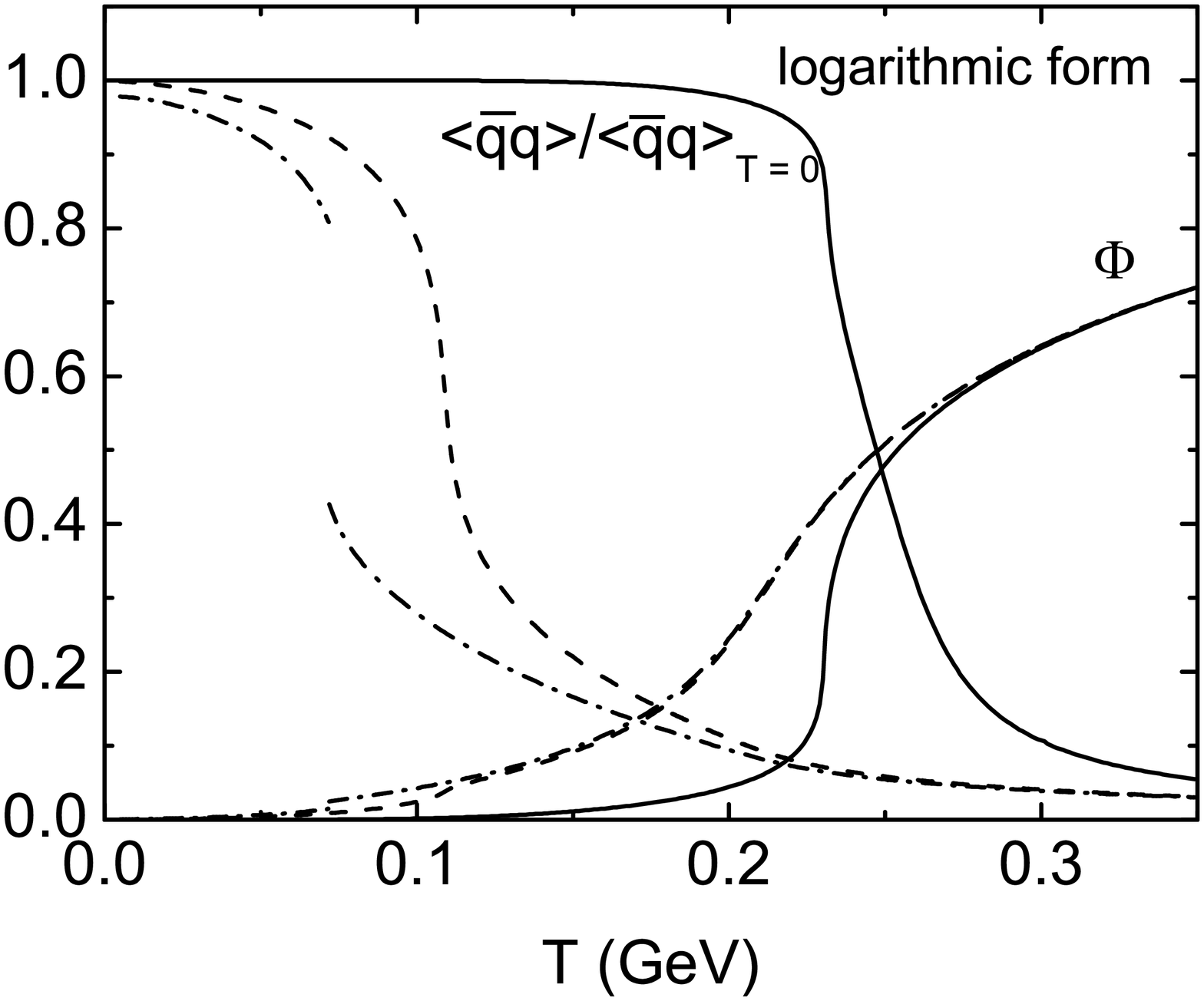, width =
7.cm} }
\caption{ Temperature dependence of the chiral condensate
and Polyakov loop potential with the new set of parameters within
the PNJL model. Solid, dashed and dot-dashed lines are calculated
for $\mu =$0, $\mu =\mu_{CEP}$ and $\mu >\mu_{CEP}$, respectively.
}
 \label{order}
\end{figure}

 To define the crossover transition line, the chiral condensate $<\bar{q}q>$
and the Polyakov loop $\Phi$  were used as
the order parameters. As shown in Fig. \ref{order},  these
quantities are the temperature-dependent functions and demonstrate
a quick change near the transition line which essentially depends
on the chemical potential $\mu$. The position of this line is
defined by local maximum of $d<\bar{q}q>/dT=$0 and $d\Phi/dT=$0.
To find the first order transition line, it is convenient to
introduce the baryon number susceptibility  $\chi_q = \dfrac{d
n_q}{d\mu}\vert_{T = const}$. The first order phase transition
ends just at point where $\chi_q$ has a pronounced maximum and
this point is called critical endpoint (CEP) where the phase 
transition of the second order. At $T \geq T_{\rm
CEP}$ the baryon number susceptibility has a sharp rise and it 
can be considered as the presence of an ideal gas of weakly 
interacting quarks.

\begin{figure}[thb]
\centerline{
\psfig{file = 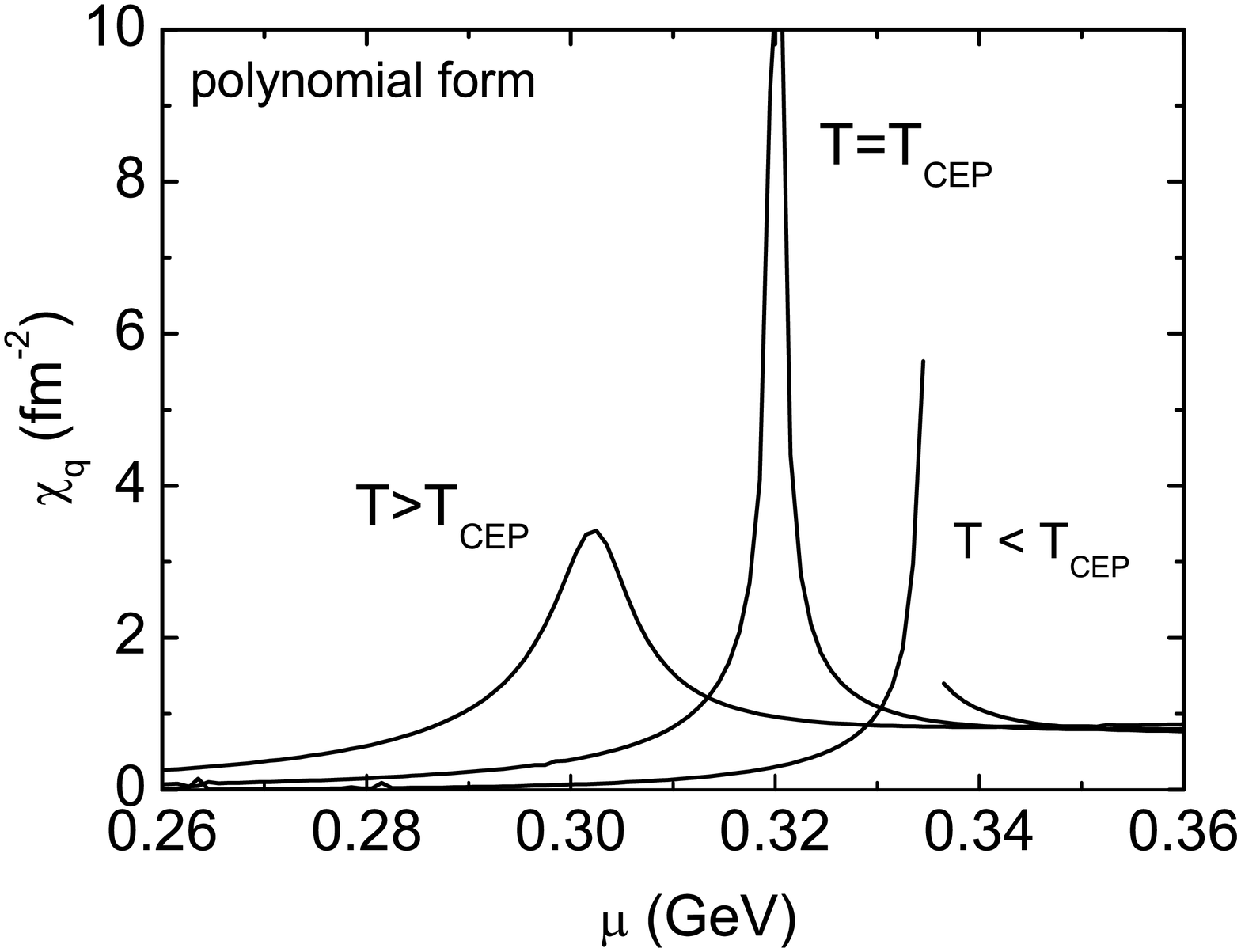, width=7.cm}\hspace*{-10mm}
\psfig{file = 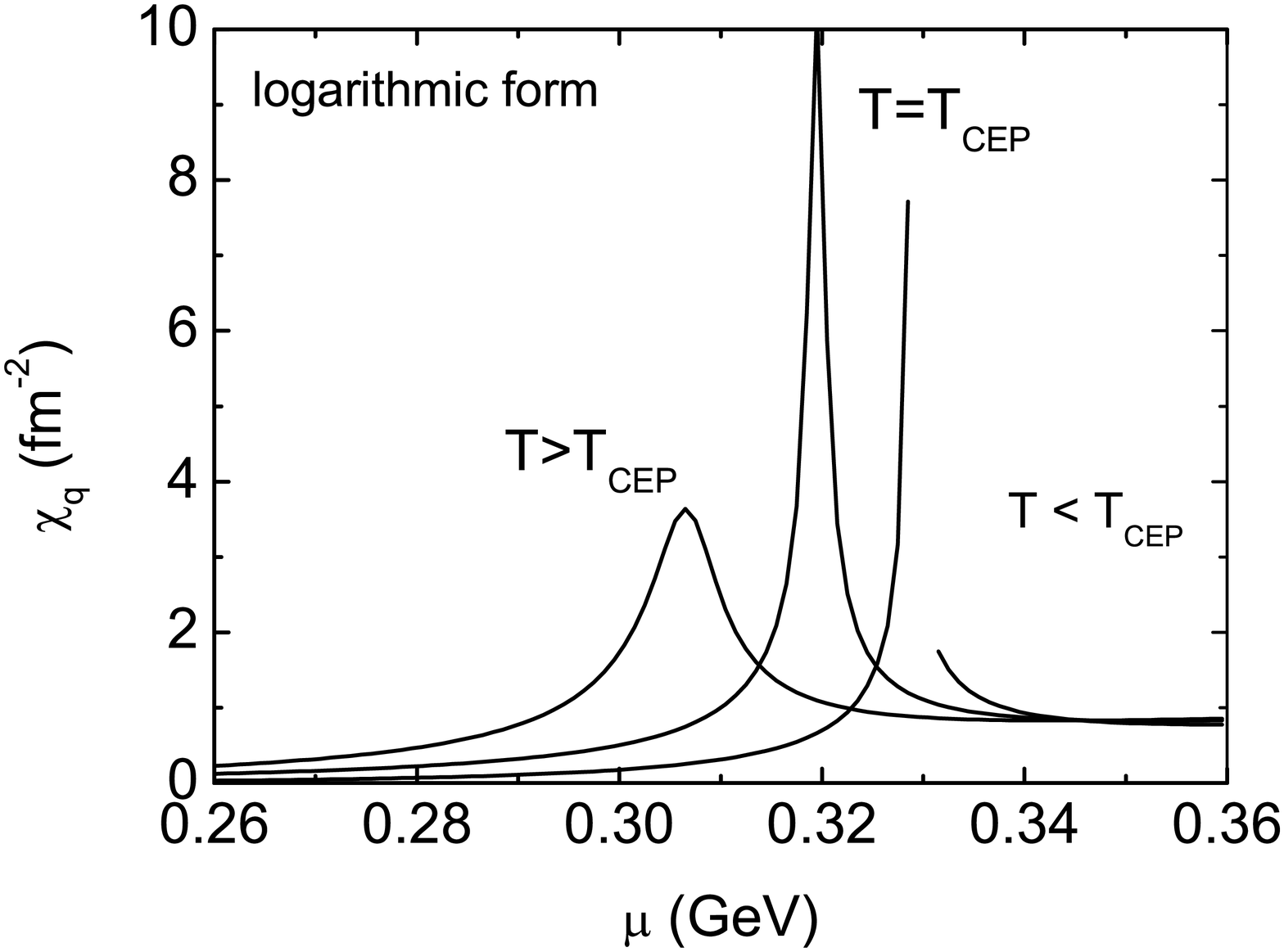, width=7.cm} }
\caption{Baryon number susceptibility around the critical endpoint
 as a function of the chemical potential for polynomial (left panel) and
 logarithmic (right panel) form of potential.
}
\label{qdens}
\end{figure}

The behavior of the baryon number susceptibility $\chi_q$ as a
function of the chemical potential for three different
temperatures around the CEP is presented in Fig.\ref{qdens}. For
$T<T_{CEP}, \mu>\mu_{CEP}$ we have a phase transition of the first
order with clear discontinuity; for $T=T_{CEP}$ the susceptibility
$\chi_q$ diverges at $\mu=\mu_{CEP}$; for $T>T_{CEP}$ the
discontinuity at the transition line disappears and we observe
crossover type of the phase transition. The polynomial and
logarithmic approximations of the Polyakov loop predict very similar
results. Similar  behavior exhibits also the specific heat
$c_v$.

As is seen from Fig. \ref{qdens}, both models show the CEP at the
temperature $T_{CEP}=T_\chi$ below which the chiral phase
transition is of the first order. At this point
($T_{CEP},\mu_{CEP}$) the phase transition changes from the first
order to crossover~\cite{costa2,kashiwa,fukushima}. At this point
the second order transition is present.

\begin{figure}[thb]
\centerline{ \psfig{file = 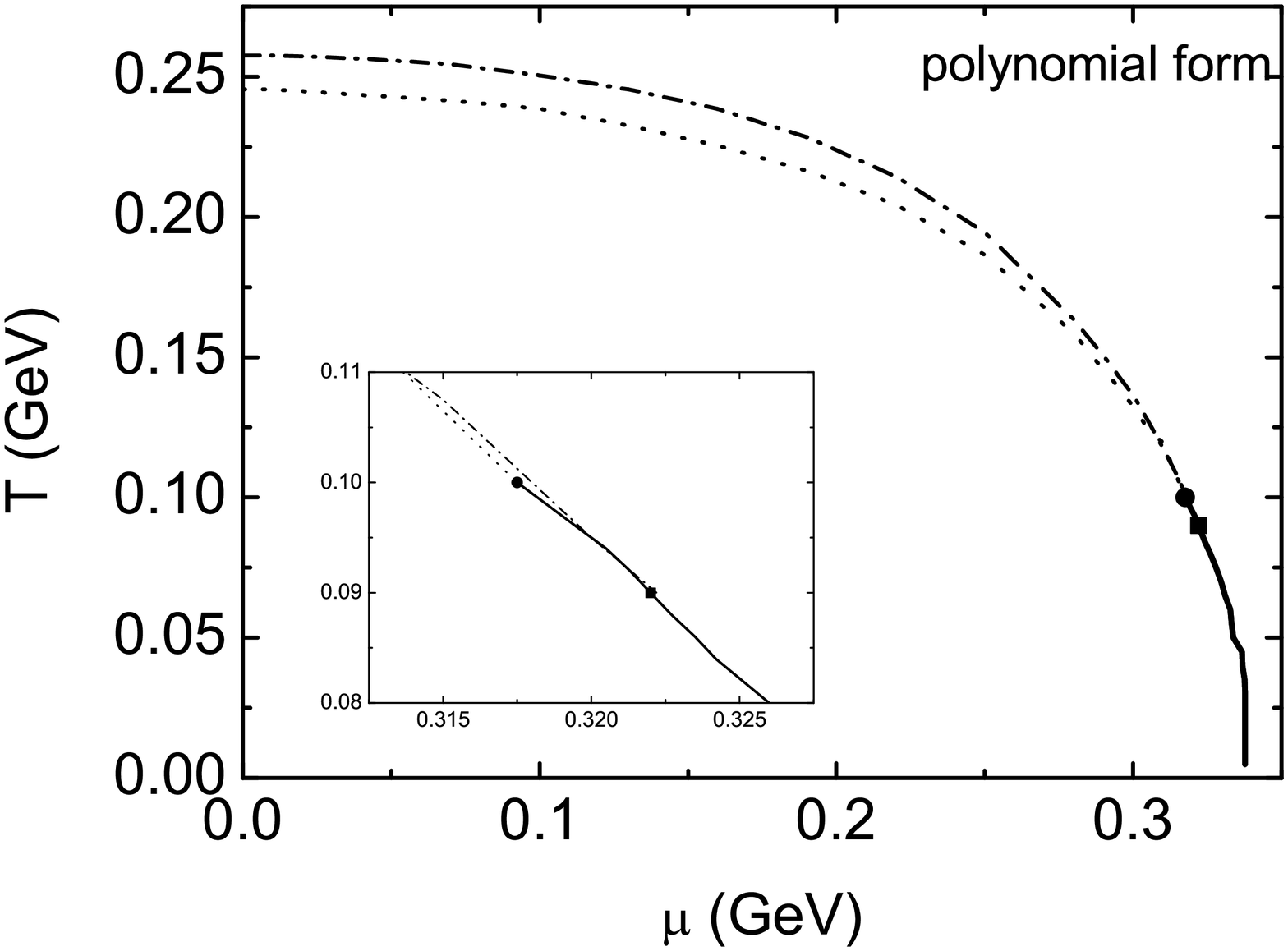, width = 7.
cm}\hspace*{-10mm} \psfig{file = 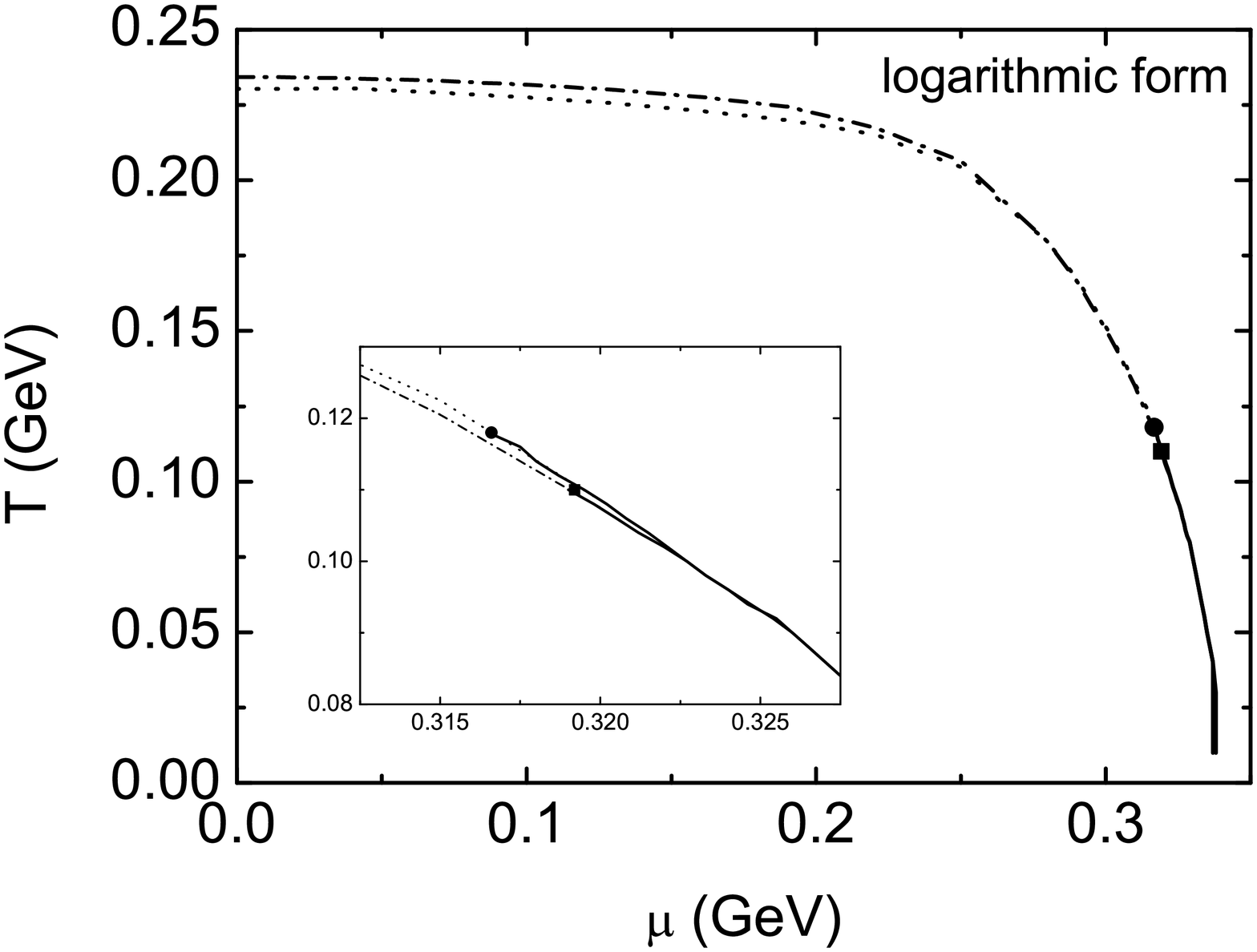, width = 7.
cm} } \caption{Phase diagrams of the PNJL model with polynomial
(left panel)  and logarithmic (right panel) forms of the
potential. Solid lines correspond to the first order phase
transition, dotted and dot-dashed lines are crossover for new and
old lattice data, respectively. The phase region near the critical
endpoints is zoomed in the insert. }
 \label{phasediag}
\end{figure}

The phase  diagram in the $(T,\mu)$ plane is presented 
in Fig. \ref{phasediag}.
 Within the PNJL model the positions of the critical endpoints
$(T_{CEP},\mu_{CEP})$ are (0.118,0.3166), (0.11,0.3192) for the logarithmic
 form and (0.10,0.3175),(0.09,0.322) for the polynomial
form, where the first pair of numbers correspond to the new data
set and the second one is for the old data set (in GeV). As was noted in
Ref. \cite{costa2}, critical properties of observables are
significantly influenced by the chosen parameter set and
regularization procedure. As follows from Fig. \ref{phasediag},
the substitution of the new basic lattice data with using the
polynomial and the logarithmic forms for the $\mathcal{U}$
approximation influences more significantly the chemical potential
of the critical endpoint
 $\mu_{CEP}$ rather than its temperature $T_{CEP}$.

\section{Summary and conclusions}

We have considered the PNJL ($N_c=3,N_f=$2) model and investigated its
phase structure at finite $T$ and $\mu$. Two different sets of
parameters based on the old and new lattice data for the pure
gluon sector were used and for each of these sets the two
different parameterizations of the effective potential
$\mathcal{U}(\Phi,\bar{\Phi},T)$ - polynomial and logarithmic -
were applied. The thermodynamics for all the developed versions of
the PNJL model was studied and compared with the available lattice
data. Consideration of different thermodynamic observables like
pressure and energy density, their $T$ and $\mu$ behavior as well
as the quark number density serves as an important probe of the
model. We found that in spite of a noticeable disagreement between
the old and new original lattice data, the effective gluon
potentials $\mathcal{U}$ are quite close to each other and a
larger difference is due to the form of their approximation:
the logarithmic form predicts a more distinct and narrower minimum at
high $T$.

The model qualitatively reproduces both $\pi$ and $\sigma$ meson
properties in hot, dense quark matter and the rich and complicated
phase structure of this medium providing information on the order
of phase transitions and the position of critical points. This
information depends  stronger on the form of the effective
potential rather than on the used lattice data set.

Unfortunately, the position of the calculated CEP in the $(T,\mu)$
plane is still far from the predictions of lattice QCD and
empirical analysis. Further elaboration of the presented model is
needed. In particular, the inclusion of entanglement interactions
between quark and gauge degrees of freedom in addition to the
covariant derivative in the original PNJL model~\cite{SSKY11} and
the incorporation of explicit diquark degrees of
freedom~\cite{RRW07} are of great interest. Both
modifications~\cite{SSKY11,RRW07} reproduce lattice data at
$\mu\geq 0$ better than the original PNJL model and influence the
position and the nature of the critical endpoint in the $(T,\mu)$
phase diagram. Moreover, with the use of the logarithmic form of
the effective potential, these models result in the appearance 
of new phases.

One should note that  we are restricted to the case 
without diquark correlations and thus 
possible color superconducting phases at low $T$ and high $\mu$
are ignored. It is attractive also to  include into consideration 
the color superconducting phases and nonlocality of the 
interaction~\cite{GomezDumm:2005hy} as well as effects beyond 
the meanfield~\cite{Blaschke:2007np,RBBV11}.

\section*{Acknowledgments}

We are grateful to D. Blaschke and P. Costa for useful comments
and constructive suggestions.
V.T. acknowledges financial support from the Helmholtz
International Center (HIC) for FAIR within the LOEWE program. The
work of Yu. K. was supported by RFBR grant No. 09-01-00770a.

\end{document}